# Enhanced solar evaporation using a photo-thermal umbrella: towards zero liquid discharge wastewater management

Akanksha K. Menon[1$], Iwan Haechler[1$], Sumanjeet Kaur[1], Sean Lubner[1], Ravi S. Prasher[1,2*]

[1]Energy Storage & Distributed Resources Division, Lawrence Berkeley National Laboratory, Berkeley, CA 94720, USA

[2]Department of Mechanical Engineering, University of California, Berkeley, CA 94720, USA

* e-mail: rsprasher@lbl.gov

[$] Equal contributions

## Abstract

Rising water demands and depleting freshwater resources have brought desalination and wastewater treatment technologies to the forefront. For sustainable water management, there is a global push towards Zero Liquid Discharge (ZLD) with the goal to maximize water recovery for reuse, and to produce solid waste that lowers the environmental impact of wastewater disposal. Evaporation ponds harvest solar energy as heat for ZLD, but require large land areas due to low evaporation rates. Here, we demonstrate a passive and non-contact approach to enhance evaporation by more than 100% using a photo-thermal umbrella. By converting sunlight into only mid-infrared radiation where water is strongly absorbing, efficient utilization of solar energy and heat localization at the water's surface through radiative coupling are achieved. The device's non-contact nature makes it uniquely suited to treat a wide range of wastewater, and the use of commercially available materials enables a potentially low cost and scalable technology for the sustainable disposal of wastewater, with the added benefit of salt recovery.

The World Economic Forum recognizes water crises as a major global risk that has arisen from the depletion of natural freshwater resources due to agricultural, industrial and municipal use, while generating vast amounts of wastewater.[1] This poses a sustainability challenge that currently threatens four billion people worldwide and is expected to become more severe with population growth and economic development.[2,3] Desalination of seawater and inland brackish water has emerged as a solution to meet increasing water demand. However, desalination plants produce concentrated brine as a byproduct, disposal of which is detrimental to land vegetation and the aquatic ecosystem, thereby having a significant environmental impact.[4,5] Thus, there has been a push towards maximizing water recovery for reuse from industrial wastewater and desalination brine to achieve Zero Liquid Discharge (ZLD) such that the final waste is a solid. At present, ZLD involves a series of treatment processes that *(i)* reduce the volume of wastewater using membrane-based systems or thermal brine concentrators, and *(ii)* reduce the concentrated brine to solid waste using a crystallizer or an evaporation pond.[6]

The choice for this final step of ZLD depends upon various factors including concentrated brine volume and composition, energy requirement, local climate and land costs.[7,8] A crystallizer is a complex mechanical system (forced circulation evaporator) that requires high-grade heat and electricity resulting in a large energy cost and capital cost. Although crystallizers have a small site footprint, their operating costs depend heavily on the composition of wastewater, which can be exorbitant for highly scaling water (*e.g.* wastewater containing large amounts of silica).[9] In contrast, evaporation ponds rely on solar energy to passively evaporate water from any waste stream, resulting in low energy and operating costs. Currently, evaporation ponds are implemented in China, Australia, Europe (Mediterranean region), the Middle East and some areas of the U.S. where they are economically viable owing to inexpensive land and a suitable climate (arid or semi-arid).[6,9,10] Although they offer a tremendous advantage of being suitable for different wastewater streams, capital costs are high due to the large land footprint required for natural evaporation. This footprint is inversely proportional to the evaporation rate, which is inherently low due to the passive nature and inefficient use of solar energy in these ponds. To reduce the environmental impact (*i.e.,* smaller areal footprint) and capital costs, evaporation enhancement in brine disposal ponds is essential and different approaches have been implemented in this regard.[11,12]

Active methods of enhancement include Wind-Aided Intensified Evaporation (WAIV) and droplet spraying, and passive techniques include the use of solar radiation absorbing organic dyes, wetted floating fins and salt tolerant plants.[7,8,12] These methods have been shown to enhance evaporation rates by up to 35%, while WAIV has shown a 50% enhancement at high salinities but requires continuous electric pumping making it an active system. Recently, a new passive approach for solar evaporation enhancement has emerged where the emphasis is to avoid wastefully heating a large volume of water and instead perform surface heating by localizing sunlight at the water–air interface.[13,14] Various prototypes using nanomaterial-based absorbers and bio-inspired structures exploiting surface heating have since been reported, with conversion efficiencies of over 90% for these floating structures.[15-20] While these are practical for solar stills and steam generation applications, their continuous operation in high salinity wastewater (such as evaporation ponds) causes salt precipitation at the surface in contact with water, resulting in deterioration of the optical and wicking properties over time.[16] Thus, there is a need to develop *non-contact* and *passive* technologies for enhanced solar evaporation that can reduce the footprint of evaporation ponds and eliminate contamination from fouling and scaling.

Here, we demonstrate a novel surface heating approach that enhances evaporation by over a 100% under 1 sun, with the potential to increase evaporation by 160% compared to traditional evaporation ponds through thermal design. The system relies solely on radiative coupling using a photo-thermal conversion device comprising a selective solar absorber and blackbody emitter, as shown in Figure 1. Since this photo-thermal device shields the water in an evaporation pond from direct sunlight, we refer to it as a "solar umbrella." In this paper, we design a lab-scale system to experimentally demonstrate the potential of the solar umbrella for enhancing evaporation from concentrated brines. Using thermal models validated by these lab-scale experiments, we predict the performance of a large-scale evaporation pond for sustainable wastewater disposal. By enhancing the evaporation two-fold, the land required for the same volume of wastewater disposal is halved, which has significant environmental and sustainability impact.

**Concept: increasing surface temperature with radiative heat localization**

Efficient utilization of solar energy for evaporation is limited by the transparency of water at visible and near-IR wavelengths owing to its low absorption coefficient of 0.01 m[-1].[21] As a result, a large fraction (~60%) of incident solar flux results in volumetric or sensible heating of bulk water as shown in Figure 1a. Although the temperature of water increases as sunlight is absorbed, that increase is distributed in the whole volume of liquid (large thermal mass) resulting in a lower temperature at the water surface, as compared to the case where all the radiation is absorbed at the surface. As a result, the transient thermal response of traditional evaporation ponds is slow, which when coupled with the diurnal variation of solar radiation, results in low evaporation rates. At mid-IR and larger wavelengths however, the absorption coefficient of water increases by several orders of magnitude to 10[4] m[-1], and consequently, radiation is absorbed and localized close to the surface (small thermal mass). Given that evaporation is a surface phenomenon, by shifting the solar radiation to mid-IR and larger wavelengths, surface heating can be achieved and evaporation can be enhanced. As an example, a blackbody with a temperature < 150 °C emits ~99.9% of radiation at wavelengths above 2 μm that can be absorbed in a thin layer of water ~100 μm (Supplementary Information). We achieve the shifting of solar radiation to mid-IR and larger wavelengths by utilizing a photo-thermal converter comprising a selective solar absorber and a blackbody emitter shown in Figure 1b. When used as an umbrella over a water surface such as an evaporation pond, the overall efficiency ($\eta$) of the system can be expressed as:

$$\eta = \left(\frac{\dot{m}h_{fg}}{q_{solar}}\right) = \eta_1 \times \eta_2 \times \eta_3 = \left(\frac{\alpha_s q_{solar} - \varepsilon_s \sigma T_{abs}^4}{q_{solar}}\right)\left(\frac{F\varepsilon_b \sigma T_{abs}^4}{\alpha_s q_{solar} - \varepsilon_s \sigma T_{abs}^4}\right)\left(\frac{\dot{m}h_{fg}}{F\varepsilon_b \sigma T_{abs}^4}\right) \qquad (1)$$

where $\dot{m}$ is the water evaporation rate, $h_{fg}$ is the latent heat of vaporization and $q_{solar}$ is the solar flux. There are three parts to this efficiency: $\eta_1$ represents the absorber efficiency, $\eta_2$ incorporates the emitter efficiency as well as the radiative coupling between the solar umbrella and water surface, and $\eta_3$ represents the fraction of incident radiation that results in evaporation.

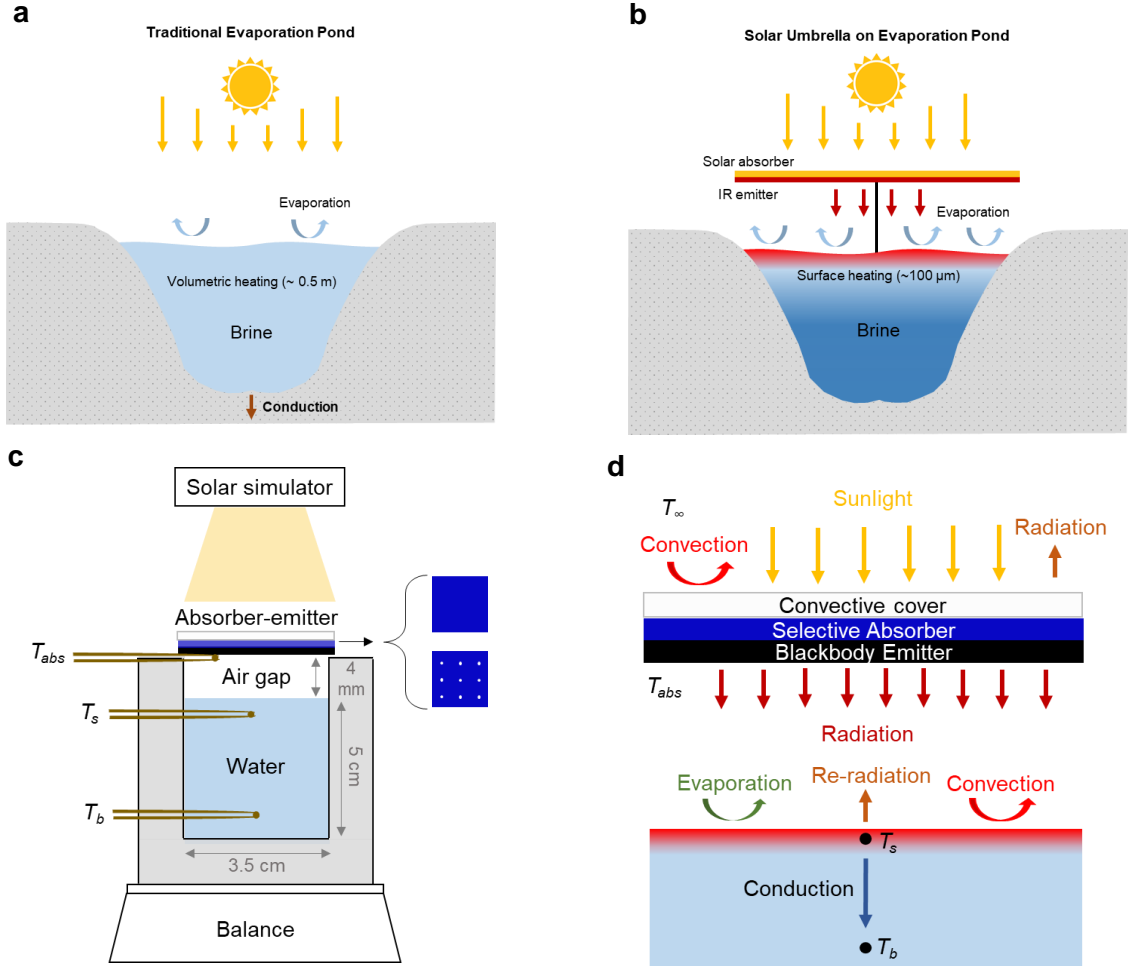

**Fig 1: Surface heating using a photo-thermal (solar) umbrella. a** Schematic of a conventional evaporation pond where incoming sunlight is volumetrically absorbed, causing a bulk water temperature increase that leads to evaporation and conduction losses to the ground. **b** Rendering of the proposed solar umbrella (spectrally selective absorber and blackbody emitter) that converts incoming sunlight into mid-IR radiation where water is strongly absorbing, thereby increasing the surface temperature and evaporation rate while the bulk remains at a lower temperature. **c** Schematic of the lab-scale prototype of an evaporation pond comprising water in an acrylic tank separated from the solar umbrella by an air gap and placed under a solar simulator to measure temperatures and mass loss due to evaporation. **d** Energy balance and modes of heat transfer for the umbrella and water, where the red region represents non-contact heat localization at the surface.

Equation (1) can be used to guide the material selection and structure of the solar umbrella. To obtain a high $\eta_1$, the top surface of the umbrella is coated with a spectrally selective solar absorber; an ideal selective absorber is characterized by a high solar absorptance, $\alpha_s$, from 0.2 – 2.5 μm and a near-zero thermal emittance, $\varepsilon_s$, at wavelengths larger than 2.5 μm.[22] This reduces radiation losses to the ambient and achieves efficient conversion of sunlight into heat at an equilibrium temperature of $T_{abs}$. To obtain a high $\eta_2$, the bottom surface of the umbrella is coated with a black emitter characterized by a high IR emittance, $\varepsilon_b$, which is placed above the water surface with a large view factor, $F$; $\sigma$ is the Stefan-Boltzmann constant. Finally, to obtain a high $\eta_3$, thermal losses from water via conduction, convection and radiation must be minimized. However this efficiency cannot be optimized to a great extent as the surface of the water is exposed to the ambient. To test the effectiveness of the non-contact surface heating approach, we designed a lab-scale prototype that consists of two sub-systems, namely the solar umbrella and a water tank, as shown in Figure 1c. Figure 1d shows the energy balance for the overall system.

**Lab-scale experimental demonstration**

In the lab-scale prototype, the solar umbrella is comprised of an aluminum substrate with a selective absorber coating on top (TiNOX, Almeco)[23] and a black paint coating (Zynolyte Hi-Temp Paint, Aervoe) on the bottom. The absorber has a solar absorptance of 0.95 and thermal emittance of 0.04 at IR wavelengths that suppresses radiation losses, while the black paint has an emittance of 0.94 at IR wavelengths (Supplementary Information). Since the substrate is thin (0.5 mm) and metallic, temperatures of the absorber and emitter are equal and remain isothermal over the course of the measurements. The tank is made of acrylic and is filled with pure water, with the solar umbrella positioned 4 mm above the water surface. A piece of transparent plastic with a transmittance of 0.91 is used as a convective shield to reduce losses from the top surface of the umbrella as shown in Figure 1c. A solar simulator was used to provide a flux of 1 sun (1000 W m⁻²) on the top surface of the umbrella. The temperatures of the solar umbrella ($T_{abs}$), water surface ($T_s$, thermocouple placed 2 mm below water surface) and bulk water ($T_b$, thermocouple placed 40 mm below water surface) were monitored over the course of the experiment. Under illumination, the temperature of the solar umbrella increases rapidly and reaches a steady-state value of $T_{abs} = 70$ °C in under three minutes (Supplementary Information). The corresponding increase in temperature of pure water due to radiative heating from the hot emitter is shown in Figure 2a. The water surface temperature steadily rises from 22.5 °C to 40 °C, while the bulk water increases only by 2.5 °C in one hour, which confirms the surface heating effect. To mimic concentrated wastewater in evaporation ponds, the experiment was repeated with a saturated common salt solution (25 wt. % NaCl). Similar trends were observed for the brine, with a 20 °C temperature rise at the surface over one hour (~2 °C higher than with pure water), thereby demonstrating radiative heat localization using the solar umbrella. The presence of salts such as NaCl strongly absorber near-IR radiation which increases the brine temperature compared to pure water. To facilitate a comparison with traditional evaporation ponds where water is illuminated directly by sunlight, the same setup was used without the solar umbrella, and corresponding temperatures are shown in Figure 2b. Due to the small absorption coefficient of water at solar wavelengths, the solar flux is volumetrically absorbed resulting in a 5 °C temperature rise throughout the water in one hour.

Given that evaporation varies with surface temperature, the mass change due to evaporation was measured for each case at an ambient temperature, $T_\infty = 25$ °C and a relative humidity of 50%. Figure 2c shows the evaporation rates under one sun after subtracting the evaporation in an otherwise identical but dark environment (0.07 kg m$^{-2}$ h$^{-1}$). Without the solar umbrella, pure water absorbs sunlight volumetrically that increases evaporation to 0.3 kg m$^{-2}$ h$^{-1}$ compared to the dark case. With the solar umbrella however, the evaporation increases to 0.62 kg m$^{-2}$ h$^{-1}$ as a higher surface temperature is achieved; this represents a 2-fold enhancement under the same input solar flux. This evaporation rate and surface temperature of 40 °C under one sun are comparable to other surface heating approaches in literature that use self-supporting floating structures (single layer devices without an insulating foam), and the device demonstrated here has the added advantage of being non-contact.[24-26] The saturated NaCl brine showed an evaporation rate of 0.49 kg m$^{-2}$ h$^{-1}$, which represents a 21% reduction compared to pure water. This is expected as vapor pressure decreases with salinity but is partially compensated by the higher temperature of brine compared to pure water. Figure 2d compares the evaporation rate over a longer period of time, resulting in precipitation and recovery of salts which confirms that the solar umbrella can be used to concentrate wastewater and achieve zero liquid discharge. The effect of the photo-thermal umbrella on water evaporation was also measured under low optical concentrations of 2, 3 and 5 suns. As expected, higher evaporation rates up to 1.6 kg m$^{-2}$ h$^{-1}$ were observed at higher input fluxes (Supplementary Information). We note that these low optical concentrations can be achieved with passive non-tracking concentrators.[27]

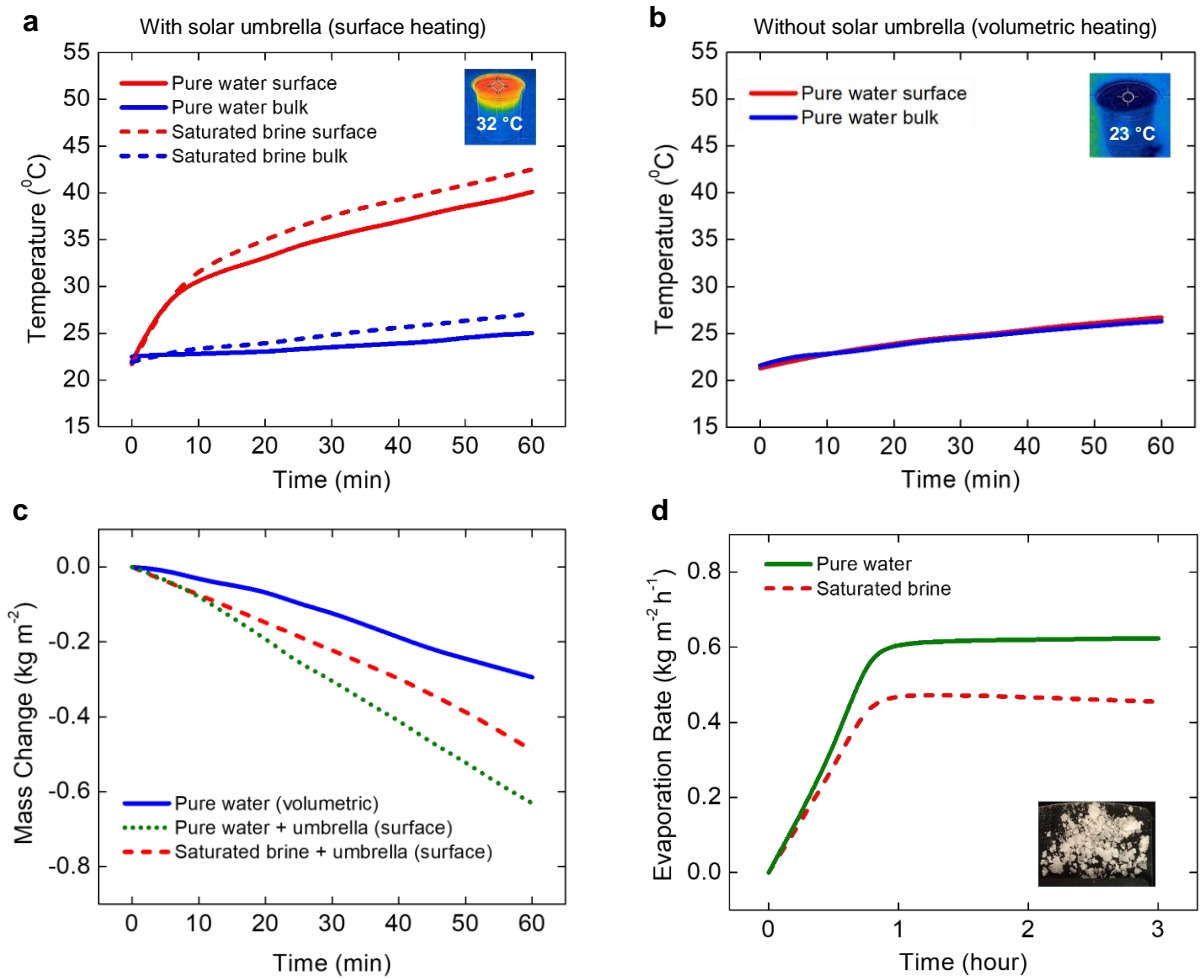

**Fig. 2: Lab-scale experimental results. a** Surface and bulk temperature of pure water (solid line) and 25 wt. % NaCl brine (dashed line) over one hour with the solar umbrella showing localized heating at the surface. **b** Surface and bulk temperatures of pure water without the solar umbrella showing volumetric heating. Inset images are captured using an IR camera after 10 minutes of illumination for visualization of surface vs. volumetric heating with and without the solar umbrella, respectively (IR images are for qualitative observation only; temperature measurements were made with thermocouples as shown in Fig. 1c). **c** Mass change due to evaporation for pure water under direct illumination (volumetric heating) compared to evaporation of pure water and NaCl brine with the solar umbrella (surface heating). **d** Evaporation rate over a three-hour period with the solar umbrella for pure water (solid line) and 25 wt. % brine (dashed line) showing a 21% reduction due to salinity. The inset shows precipitated salt crystals from the walls of the water tank owing to evaporation of water from brine to achieve ZLD. All experiments are conducted under one sun (1000 W m$^{-2}$) at an ambient temperature of 25 °C and relative humidity of 50%.

For the case of pure water, these experimentally measured values are used to calculate the sub-system efficiencies from Equation (1): $\eta_1 = 91\%$ $\eta_2 = 74\%$ and $\eta_3 = 63\%$, resulting in an overall solar-thermal evaporation efficiency of 43% under one sun. This compares favorably with reported single-layer floating evaporation structures[20,28] and has the added benefit of being non-contact thus eliminating contamination from the wastewater. The absorber efficiency of 91% can be explained in terms of losses due to reflection and thermal radiation as evidenced from the optical property measurements (Supplementary Information). The emitter efficiency of 74% is due to an emittance and view factor less than unity ($\varepsilon_b = 0.94$ and $F = 0.9$) in the lab-scale prototype, as well as convection losses from the umbrella even with a cover. These convective losses can be minimized with an evacuated absorber that is commonly used in domestic solar hot water heaters. Finally, an evaporation efficiency of 63% indicates that there are thermal losses via conduction, convection and radiation in the system that compete with and reduce the energy available for evaporation, as discussed in the next section. Nevertheless, a solar-thermal evaporation efficiency of 43% is achieved in this first demonstration of non-contact surface heating. In comparison, the solar-thermal efficiency of volumetrically heated water under the same solar flux is ~20% based on the experimentally measured evaporation rate (see Figure 2c) and consistent with literature,[18] thus confirming the superior performance of surface-based heating for solar evaporation.

**Thermal model validation**

To estimate heat losses in the lab-scale prototype and to gain insight into improving performance, a thermal model was developed using COMSOL Multiphysics. Relevant boundary conditions and material properties of the experimental setup were used as inputs (Supplementary Information). The evaporation was modeled as a boundary heat flux:

$$q_{evap} = h_{evap}(T_s - T_\infty) = \dot{m}h_{fg} \qquad (2)$$

where $q_{evap}$ is the heat loss due to evaporation characterized by an effective heat transfer coefficient, $h_{evap}$ (which can in general be a function of temperature). Using the experimentally measured $\dot{m}$ and $T_s$ values obtained at different optical concentrations (see Supplementary Information), an average value of $h_{evap}$ = 28 W m$^{-2}$ K$^{-1}$ is extracted which matches floating structure prototypes.[13] The low value suggests a diffusion-limited process[19,29] that is likely to improve when moving from lab scale to deployment scale due to wind.

The simulated cross-sectional temperature profile of water in the acrylic tank at steady-state is shown in Figure 3a. Heat localization results in a higher temperature of 42 °C at the surface, while the bulk is close to ambient at 25 °C. Figure 3b shows the transient temperature over an hour, which matches the experimentally measured values, thereby confirming the validity of the thermal model. From this analysis, conduction from the surface into underlying bulk water was calculated to be 21%, convection from the walls of the acrylic tank were 6% and radiation from the acrylic walls were 11%. This leaves 62% of the incident energy on the water surface for evaporation, which is in good agreement with $\eta_3$ obtained from the experimentally measured mass change due to evaporation. We note here that due to the small size of the lab-scale prototype, side walls of the acrylic tank contribute to a 17% parasitic loss implying that higher evaporation efficiencies are achievable in a larger system.

Next, the validated thermal model is used to make performance predictions under different conditions, as shown in Figures 3c and 3d. For instance, on cloudy or winter days, the solar flux can be as low as 300 W m$^{-2}$ but the umbrella can still reach 45 °C and heat the water surface to 30 °C from an initial temperature of 22 °C to enhance evaporation. Furthermore, by eliminating convective losses from the umbrella surfaces (*e.g.* evacuated absorber), the steady-state temperature of the solar umbrella can exceed 100 °C under one sun as shown in Figure 3c. This in turn increases the blackbody emission to the water surface, resulting in a temperature rise to over 50 °C as shown in Figure 3d. Given that the vapor pressure of water increases by 70% from 40 to 50 °C, much higher efficiencies and evaporation rates can be obtained using an evacuated umbrella.

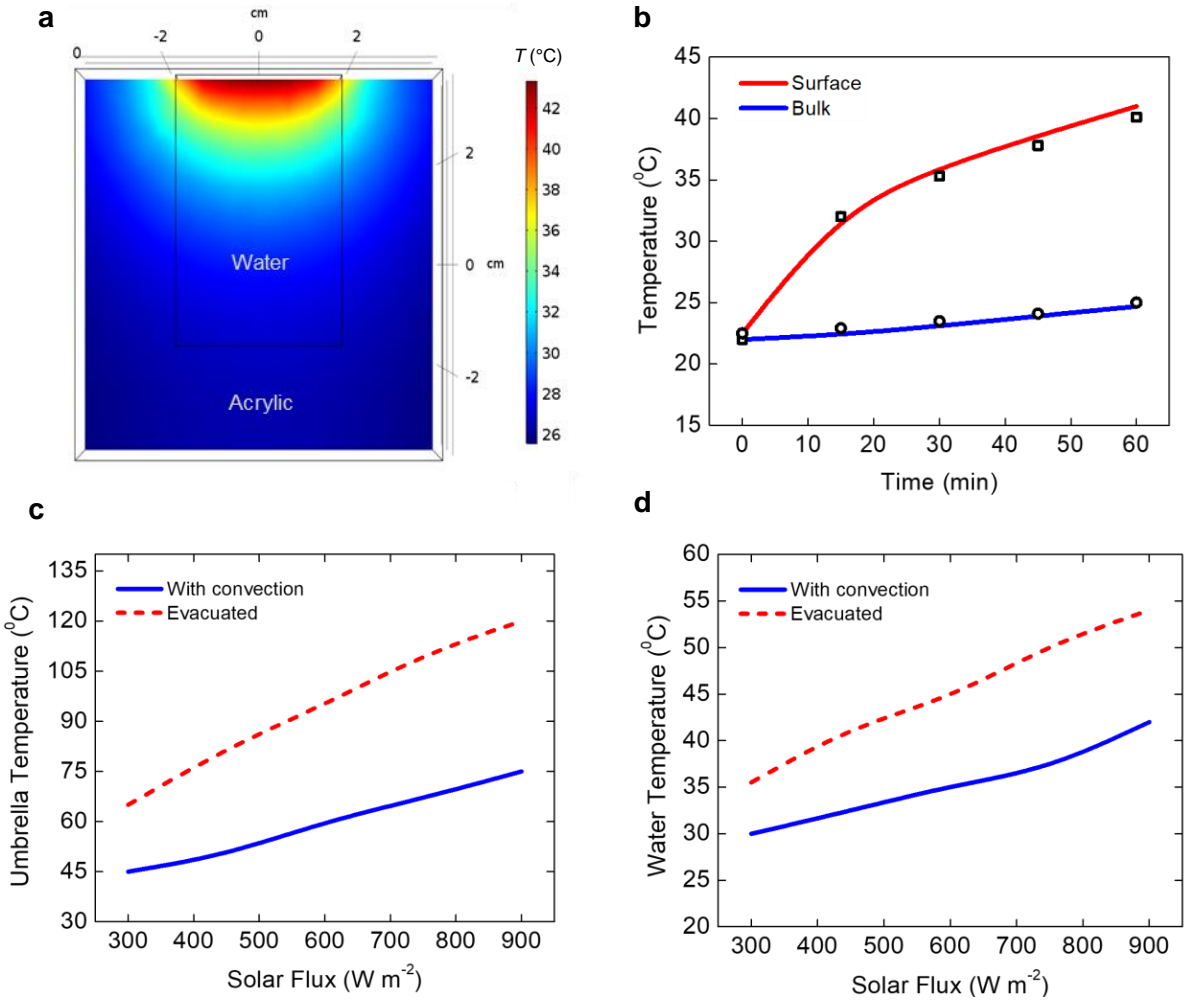

**Fig. 3: Thermal model validation and performance prediction for Solar Umbrella. a** Cross-sectional temperature profile of water in an acrylic tank with solar umbrella modeled using COMSOL that shows localized heating to a steady state surface temperature of 42 °C due to a large absorption coefficient (10⁴ m⁻¹) at mid-IR wavelengths. **b** Water surface and bulk temperatures simulated over one hour (solid lines) and overlaid with experimentally measured temperatures (open symbols) that confirms the validity of the thermal model. **c** Simulated temperature of the solar umbrella at various solar fluxes indicate that convection losses limit the performance, and **d** the corresponding simulated water surface temperature predictions show that even at low solar fluxes, the water surface temperature increases and can reach 55 °C if an evacuated system is used.

**Performance prediction for evaporation ponds**

Based on the experimentally validated thermal model, the performance of the solar umbrella for real evaporation ponds can be predicted accounting for the diurnal variation of sunlight. The diurnal variation of solar radiation causes a time delay of a few hours between the highest solar intensity and maximum brine temperature in traditional evaporation ponds as sunlight is volumetrically absorbed and used for the sensible heating of water. As a result, the transient response is slow and the evaporation rate is small, thereby requiring large land areas to dispose wastewater. In evaporation ponds having a white salt precipitate layer at the bottom, this is exacerbated due to further losses by reflection and conduction to the ground,[30] resulting in even lower solar-thermal evaporation efficiencies. One commonly used approach to address this is the addition of colored organic dyes (*e.g.*, naphthol green and methylene blue) that increase the solar absorption of water during the day. This has been shown to create a ~3 °C increase in brine surface temperature which enhances the evaporation by up to 35%, but at night these dyes lead to lower surface temperatures.[30,31] The solar umbrella presented in this work can serve as an alternative to these colored dyes and eliminate the time lag through localizing heat at the evaporation surface. To investigate the effectiveness of this approach, the aforementioned validated COMSOL model was modified to mimic a typical evaporation pond depth[7,30] of 0.5 m at an ambient temperature of 20 °C and relative humidity of 50%. The large-scale system surpasses limitations of the lab-scale prototype as the view factor between the umbrella and pond surface is unity (for a real evaporation pond with a large surface area, all the radiation emitted by the solar umbrella reaches the water surface so $F = 1$ in this case) and parasitic losses (convection and radiation from sides of the acrylic tank) are eliminated. Under 1 sun, the solar umbrella temperature reaches 70 °C, resulting in a thermal emission of ~750 W m$^{-2}$ to the surface of the pond. The resulting surface temperature (accounting for natural convection, evaporation and radiation) for a complete 24-hour cycle with sunlight during the first eight-hours, is shown in Figure 4a. From these temperatures, the time dependent evaporation rate is predicted using:

$$\dot{m} = h_m(\rho_s - \rho_\infty)$$

where $h_m$ is the convective mass transfer coefficient for evaporation, $\rho_s$ is the water vapor density at the surface and $\rho_\infty$ is the corresponding density in ambient air. For fixed ambient conditions, $\rho_\infty$ is known and with an average value of $h_m = 0.0042$ m s$^{-1}$ (based on the lab-scale experiments and comparable with data on evaporation from a stagnant water surface[32]), an increase in the surface temperature from 20 to 40 °C in Figure 4a results in a three-fold increase in $\rho_s$. This results in a higher evaporation rate as shown in Figure 4a for pure water, where the daily evaporation rate is obtained by integrating the time and temperature-dependent mass flux yielding ~9.8 kg m⁻² day⁻¹. To account for the salinity of brines, the evaporation is reduced by 21% to 7.8 kg m⁻² day⁻¹. We note that the 21% reduction is based on the experimentally measured evaporation rates for pure water and saturated NaCl in the lab-scale prototype and is not a constant value. However, it has been reported that for a salinity of 25 wt. % and at similar ambient conditions to the one modeled here, this value varies between 18 – 22%;[33] thus a fixed reduction of 21% is valid for this analysis. We also note that in real ponds $h_m$ can be higher due to the presence of wind, and other factors such as low humidity in arid regions would also increase evaporation. As such, our model predictions under natural convection and relative humidity of 50% provide a conservative estimate of the performance enhancement.

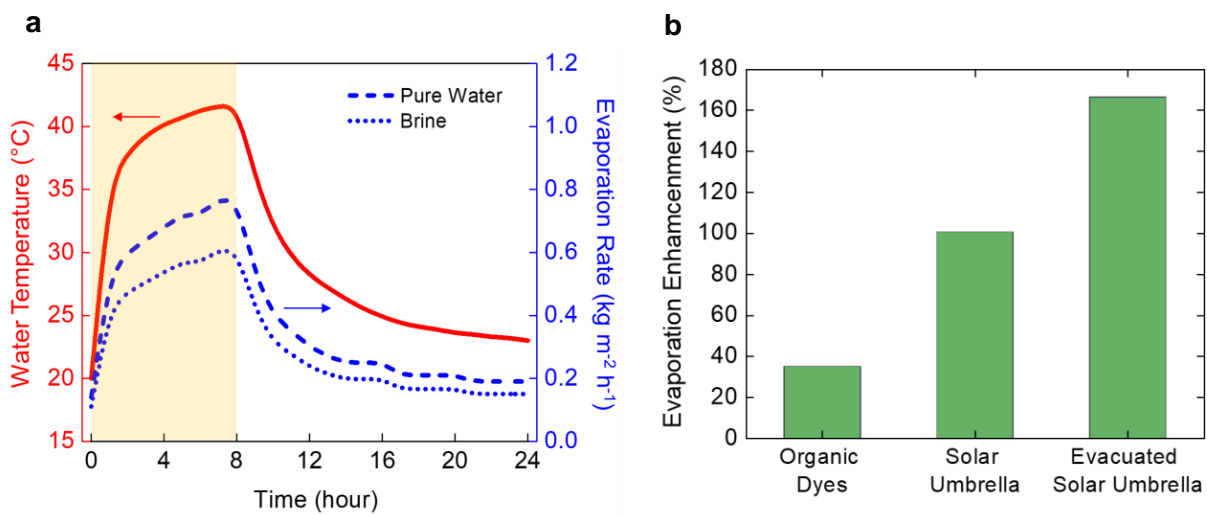

**Fig. 4: Simulated performance of evaporation ponds for ZLD. a** Simulation of water surface temperature and evaporation rate over a diurnal cycle; the shaded region represents sunlight incident on the solar umbrella, resulting in heat localization at the water surface and a corresponding rise in temperature. The evaporation rate is calculated at each temperature, and the brine evaporation rate under these conditions is obtained as a 21% reduction from pure water evaporation. **b** Daily evaporation enhancement (relative to a traditional volumetrically heated pond under 1 sun) by passive methods including radiation absorbing dyes, solar umbrella and an evacuated solar umbrella at a relative humidity of 50% and ambient temperature of 20 °C.

The main advantage is the fast response of the solar umbrella that allows for efficient evaporation during the 8-hour illumination window, which represents over a 100% enhancement compared to volumetrically heated brines under similar ambient conditions over a diurnal cycle.[33-36] An evacuated umbrella may be used for further performance enhancement as discussed in Figure 3, and in this case, a daily evaporation of 14.3 kg m$^{-2}$ day$^{-1}$ can be obtained (or 11.3 kg m$^{-2}$ day$^{-1}$ accounting for brine salinity). This leads to a ~ 160% enhancement in the evaporation rate compared to traditional evaporation ponds, thus surpassing organic dyes and WAIV.

In summary, we report a new photo-thermal approach to improve the efficiency of solar evaporation passively without making contact with the water surface. This leverages the inherent absorption properties of water, *i.e.,* a large absorption coefficient at mid-IR wavelengths that causes heat to be localized in a thin layer (<100 µm) at the water's surface. The system comprises a commercially available selective solar absorber, and the emitter is an inexpensive spray on black paint. The solar umbrella with surface heating shows over a 100% enhancement in evaporation compared to volumetric heating that can lower the land requirement, and a 43% solar-thermal efficiency under one sun. The non-contact radiative coupling eliminates risk of fouling and contamination from wastewater streams, making this system uniquely suitable for implementation in brine disposal ponds for achieving zero liquid discharge. Furthermore, by reducing thermal losses (primarily convection from the absorber surface), high temperature vapor or steam generation can be achieved under one sun which has applications in power generation and sterilization. Another potential application of the photo-thermal umbrella is in solar stills where vapor generated can be condensed to produce drinking water for off-grid communities.

**Methods**

*Materials and characterization*: The selective absorber was obtained from Almeco (TiNOX$_{energy}$ on aluminum) and the black paint was an off-the-shelf spray paint (Zynolyte Hi-Temp Paint, Aervoe) that is stable up to temperatures of 650 °C and was cured at 350 °C for two hours prior to testing. The optical properties of spectrally selective absorber and black paint emitter were measured using an FTIR (Thermo Electron Nicolet 5700), coupled with an integrating sphere accessory (Pike Technologies Mid-IR IntegratIR). As the samples were non-transmitting, the absorptance was calculated as unity minus reflectance.

*Experimental setup*: For lab-scale prototype testing, a solar simulator (Newport, 94081A) with an optical filter for AM 1.5G spectrum is used as the solar input. A power meter and thermopile detector (Newport, 919P-030-18) were used to measure the incoming solar flux at the same location as the water tank. The tank is made out of acrylic and comprises an inner pocket (square cross-section with a side length of 35 mm and a 5 mm depth) that is coated in a reflective foil and filled with water. The inner pocket is surrounded by a 20 mm thick acrylic wall on all sides that minimizes thermal losses from the sides of the tank. Two K-type thermocouples were placed in the tank: one ~2 mm below the water surface to record the surface temperature, $T_s$, and one 40 mm below the surface to record the bulk water temperature, $T_b$. The selective absorber-black emitter was mounted on Teflon spacers and placed ~4 mm above the water surface. Another K-type thermocouple was attached to the surface of the emitter to measure its temperature, $T_{abs}$, and estimate the blackbody emissive power. A convective shield (plastic with solar transmittance of 0.91) was placed on top of the absorber to reduce convection losses from the surface. The tank was placed on a balance (A&D, GF-4000) that records the mass loss due to evaporation over time. A hygrometer (Onset, HOBO UX100) was kept next to the balance to measure ambient conditions (temperature, $T_\infty$ and relative humidity). The K-type thermocouples were connected to a data logger (Pico Technologies, USB TC-08) that records temperature as a function of time. An aperture was placed above the absorber to limit extraneous light during experimental testing. As a first step, evaporation under dark conditions was measured for 30 minutes with the simulator shutter closed. Next, the shutter was opened with the incident flux set to 1000 W m$^{-2}$, unless specified otherwise. Mass loss and temperature measurements were made for at least two hours, and values reported in the manuscript are averaged over four runs with the dark evaporation subtracted.

## Acknowledgements

This work was supported by the Laboratory Directed Research and Development Program (LDRD) at Lawrence Berkeley National Laboratory under contract # DE-AC02-05CH11231. The authors thank Sishir

Mohammed for assistance with thermal modeling. A.K.M acknowledges funding support from the ITRI-Rosenfeld Fellowship.

**Author Contributions**

**References**

1     The Global Risks Report 2018. (World Economic Forum, Geneva, 2018).

2     Grant, S. B. *et al.* Taking the "Waste" Out of "Wastewater" for Human Water Security and Ecosystem Sustainability. *Science* **337**, 681 (2012).

3     Vörösmarty, C. J. *et al.* Global threats to human water security and river biodiversity. *Nature* **467**, 555, doi:10.1038/nature09440

    https://www.nature.com/articles/nature09440#supplementary-information (2010).

4     Pinto, F. S. & Marques, R. C. Desalination projects economic feasibility: A standardization of cost determinants. *Renewable and Sustainable Energy Reviews* **78**, 904-915, doi:https://doi.org/10.1016/j.rser.2017.05.024 (2017).

5     Gude, V. G. Desalination and sustainability – An appraisal and current perspective. *Water Research* **89**, 87-106, doi:https://doi.org/10.1016/j.watres.2015.11.012 (2016).

6     Tong, T. & Elimelech, M. The Global Rise of Zero Liquid Discharge for Wastewater Management: Drivers, Technologies, and Future Directions. *Environmental Science & Technology* **50**, 6846-6855, doi:10.1021/acs.est.6b01000 (2016).

7     Morillo, J. *et al.* Comparative study of brine management technologies for desalination plants. *Desalination* **336**, 32-49, doi:https://doi.org/10.1016/j.desal.2013.12.038 (2014).

8     Giwa, A., Dufour, V., Al Marzooqi, F., Al Kaabi, M. & Hasan, S. W. Brine management methods: Recent innovations and current status. *Desalination* **407**, 1-23, doi:https://doi.org/10.1016/j.desal.2016.12.008 (2017).

9     Juby, G. *et al.* Evaluation and Selection of Available Processes for a Zero-Liquid Discharge System. Report No. DWPR No. 149, (U.S. Department of the Interior Bureau of Reclamation, 2008).

10    Mickley, M. Treatment of Concentrate. Report No. DWPR Report No. 155, (U.S. Department of the Interior Bureau of Reclamation, 2008).

11    Ahmed, M., Shayya, W. H., Hoey, D. & Al-Handaly, J. Brine Disposal from Inland Desalination Plants. *Water International* **27**, 194-201, doi:10.1080/02508060208686992 (2002).

12    Hoque, S., Alexander, T. & Gurian, P. L. Innovative Technologies Increase Evaporation Pond Efficiency. *IDA Journal of Desalination and Water Reuse* **2**, 72-78, doi:10.1179/ida.2010.2.1.72 (2010).

13    Ghasemi, H. *et al.* Solar steam generation by heat localization. *Nature Communications* **5**, 4449, doi:10.1038/ncomms5449

    https://www.nature.com/articles/ncomms5449#supplementary-information (2014).

14    Tao, P. *et al.* Solar-driven interfacial evaporation. *Nature Energy*, doi:10.1038/s41560-018-0260-7 (2018).

15    Shi, Y. *et al.* Solar Evaporator with Controlled Salt Precipitation for Zero Liquid Discharge Desalination. *Environmental Science & Technology* **52**, 11822-11830, doi:10.1021/acs.est.8b03300 (2018).

16    Ni, G. *et al.* A salt-rejecting floating solar still for low-cost desalination. *Energy & Environmental Science* **11**, 1510-1519, doi:10.1039/C8EE00220G (2018).

17    Xu, N. *et al.* Mushrooms as Efficient Solar Steam-Generation Devices. *Advanced Materials* **29**, 1606762, doi:10.1002/adma.201606762 (2017).

18    Finnerty, C., Zhang, L., Sedlak, D. L., Nelson, K. L. & Mi, B. Synthetic Graphene Oxide Leaf for Solar Desalination with Zero Liquid Discharge. *Environmental Science & Technology* **51**, 11701-11709, doi:10.1021/acs.est.7b03040 (2017).

19    Ni, G. *et al.* Steam generation under one sun enabled by a floating structure with thermal concentration. *Nature Energy* **1**, 16126, doi:10.1038/nenergy.2016.126

      https://www.nature.com/articles/nenergy2016126#supplementary-information (2016).

20    Bae, K. *et al.* Flexible thin-film black gold membranes with ultrabroadband plasmonic nanofocusing for efficient solar vapour generation. *Nature Communications* **6**, 10103, doi:10.1038/ncomms10103

      https://www.nature.com/articles/ncomms10103#supplementary-information (2015).

21    Segelstein, D. J. *The complex refractive index of water*, University of Missouri-Kansas City, (1981).

22    Cao, F., McEnaney, K., Chen, G. & Ren, Z. A review of cermet-based spectrally selective solar absorbers. *Energy & Environmental Science* **7**, 1615-1627, doi:10.1039/C3EE43825B (2014).

23    *TiNOX energy*.

24    Shi, L., Wang, Y., Zhang, L. & Wang, P. Rational design of a bi-layered reduced graphene oxide film on polystyrene foam for solar-driven interfacial water evaporation. *Journal of Materials Chemistry A* **5**, 16212-16219, doi:10.1039/C6TA09810J (2017).

25    Ye, M. *et al.* Synthesis of Black TiOx Nanoparticles by Mg Reduction of TiO2 Nanocrystals and their Application for Solar Water Evaporation. *Advanced Energy Materials* **7**, 1601811, doi:10.1002/aenm.201601811 (2016).

26    Zhang, L., Tang, B., Wu, J., Li, R. & Wang, P. Hydrophobic Light-to-Heat Conversion Membranes with Self-Healing Ability for Interfacial Solar Heating. *Advanced Materials* **27**, 4889-4894, doi:10.1002/adma.201502362 (2015).

27    Winston, R. Principles of solar concentrators of a novel design. *Solar Energy* **16**, 89-95, doi:https://doi.org/10.1016/0038-092X(74)90004-8 (1974).

28    Wang, Z. *et al.* Bio-Inspired Evaporation Through Plasmonic Film of Nanoparticles at the Air–Water Interface. *Small* **10**, 3234-3239, doi:10.1002/smll.201401071 (2014).

29    Hisatake, K., Tanaka, S. & Aizawa, Y. Evaporation rate of water in a vessel. *Journal of Applied Physics* **73**, 7395-7401, doi:10.1063/1.354031 (1993).

30    Bloch, M. R., Farkas, L. & Spiegler, K. S. Solar Evaporation of Salt Brines. *Industrial & Engineering Chemistry* **43**, 1544-1553, doi:10.1021/ie50499a025 (1951).

31    Gunaji, N. N., Keyes, C. G., New Mexico State, U. & United, S. *Disposal of brine by solar evaporation*.  (U.S. Dept. of the Interior, Office of Saline Water, 1968).

32    Marek, R. & Straub, J. Analysis of the evaporation coefficient and the condensation coefficient of water. *International Journal of Heat and Mass Transfer* **44**, 39-53, doi:https://doi.org/10.1016/S0017-9310(00)00086-7 (2001).

33    Harbeck Jr, G. E. The effect of salinity on evaporation. Report No. 272A, (1955).

34    Langbein, W. B. & Harbeck, G. E. Studies of Evaporation. *Science* **119**, 328 (1954).

35    Moore, J. & Runkles, J. R. Evaporation from Brine Solutions under Controlled Laboratory Conditions. Report No. 77, (1968).

36      Turk, L. J. Evaporation of Brine: A Field Study on the Bonneville Salt Flats, Utah. *Water Resources Research* **6**, 1209-1215, doi:10.1029/WR006i004p01209 (1970).

# Enhanced solar evaporation using a photo-thermal umbrella: towards zero liquid discharge wastewater management

Akanksha K. Menon[1$], Iwan Haechler[1$], Sumanjeet Kaur[1], Sean Lubner[1], Ravi S. Prasher[1,2*]

## Supplementary Information

### Zero Liquid Discharge Treatment Process

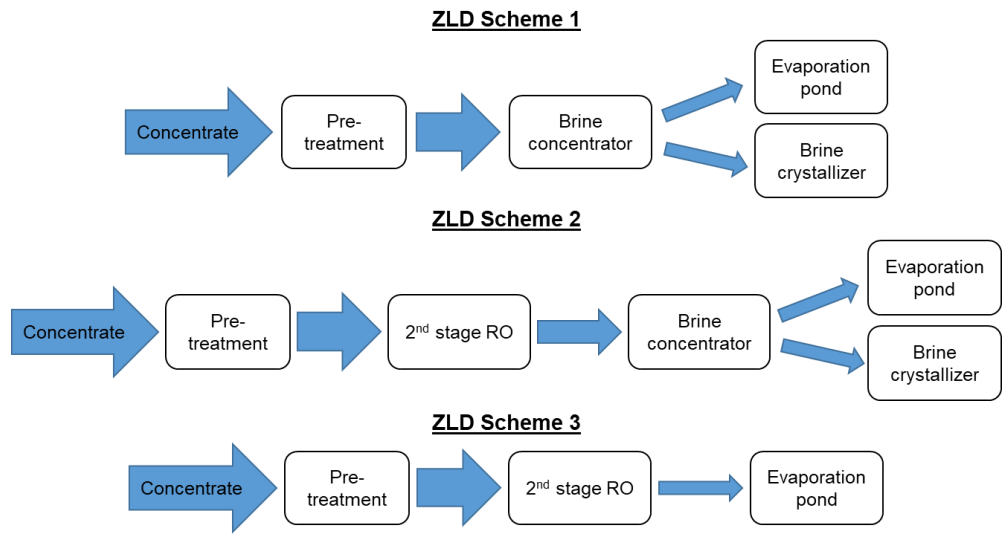

Fig. S1: Flow charts showing three different schemes for ZLD using membrane-based and thermal systems. The final step involves the use of either a brine crystallizer or an evaporation pond to extract remaining water and yields a solid waste for disposal.

### Radiative Heat Localization in Water

The absorption of electromagnetic waves (and thus light) in matter is defined by the optical properties of the material through which the wave is travelling. With Beer-Lambert-Bouguer's law, one can determine the intensity of the incident wave at a given depth $x$:

$$I(x) = I_0 * e^{-\alpha x}$$

where $I_0$ is the incident intensity on the surface of the material, $x$ is the distance from the surface and $\alpha$ is the absorption coefficient. The absorption coefficient is a function of the material properties and also depends on the wavelength of incoming radiation. For wavelengths between 250 – 800 nm (which comprises 60% of incoming sunlight), the absorption coefficient is very small ~ 0.01 m$^{-1}$ and water acts as a transparent medium. However, at wavelengths larger than 2 μm (infrared), the absorption coefficient increases by nearly 6 orders of magnitude. Consequently, the penetration depth (intensity reduced by a factor 1/e) decreases by several orders of magnitude and radiation is localized close to the surface. Predominantly IR emission can be achieved using a blackbody with a temperature, $T < 150\ °C$ (~99.9% of the thermal radiation lies at wavelengths above 2 µm), where 90-95% of radiation can be absorbed within tens of microns of water.

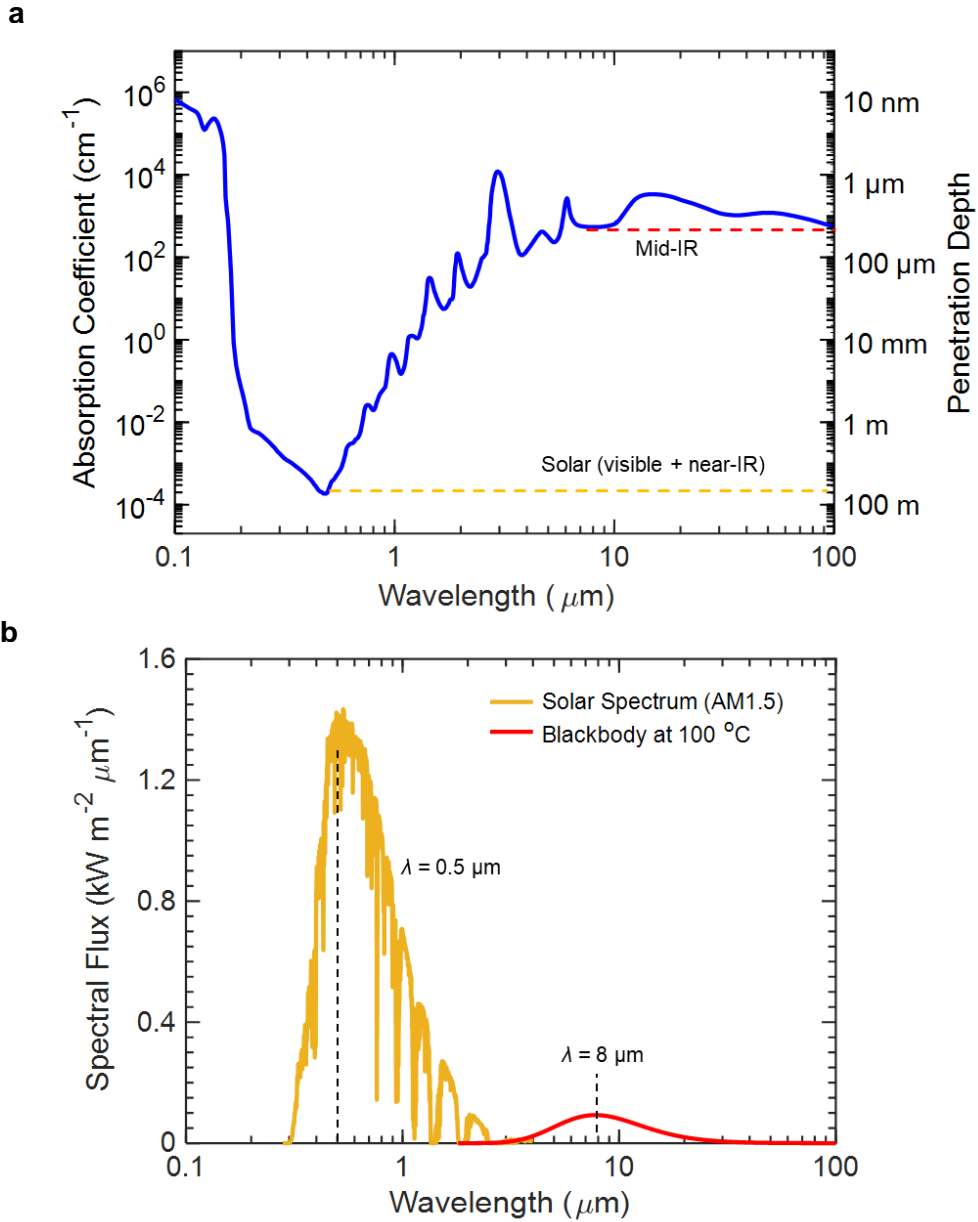

Fig. S2: (a) Absorption coefficient of electromagnetic radiation in water with penetration depth of 30 m in visible wavelengths and 20 µm at mid-IR wavelengths. (b) Blackbody emissive power of the sun ($T = 5505\ °C$) that emits in the visible and near-IR, and a blackbody at 100 °C that emits in mid-IR where water is strongly absorbing.

## Optical Characterization

The optical properties of the selective solar absorber and black paint emitter were measured using an FTIR (Thermo Electron Nicolet 5700), coupled with an integrating sphere accessory (Pike Technologies Mid-IR IntegratIR). As the samples were non-transmitting, the absorptance was calculated as unity minus reflectance.

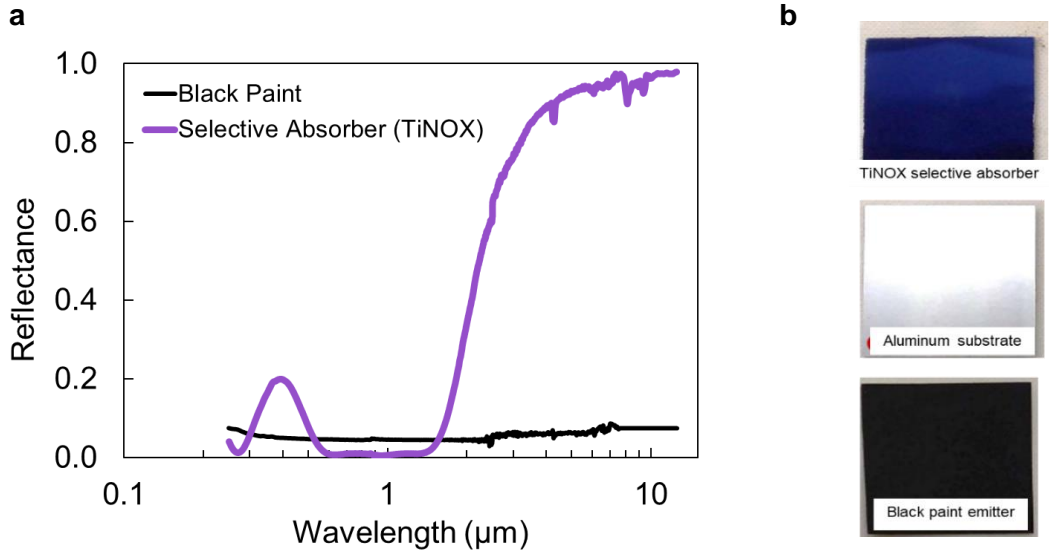

Fig. S3: (a) Reflectance of the selective absorber (TiNOX$_{energy}$, Almeco) and black paint emitter measured using an FTIR. (b) Structure of the solar umbrella comprising a selective solar absorber coated on the top surface of an aluminum substrate and a black paint sprayed on to the bottom surface.

## Lab-Scale Setup

For lab-scale prototype testing, a solar simulator (Newport, 94081A) with an optical filter for AM 1.5G spectrum is used as the solar input. A power meter and thermopile detector (Newport, 919P-030-18) were used to measure the incoming solar flux at the same location as the water tank. The tank is made out of acrylic and comprises an inner pocket (square cross-section with a side length of 35 mm and a 5 mm depth) that is coated in a reflective foil and filled with water (to reflect radiation on the tank walls into water which provides a better representation of a large water body absorbing radiation). The inner pocket is surrounded by a 20 mm thick acrylic wall on all sides that minimizes thermal losses from the sides of the tank. Two K-type thermocouples were placed in the tank: one ~2 mm below the water surface to record the surface temperature, $T_s$, and one at the bottom of the inner pocket to record the bulk water temperature, $T_b$. The selective absorber-black emitter was mounted on Teflon spacers and placed ~3-4 mm above the water surface. In this scenario, the air gap between the hot emitter and cooler water surface acts as an insulting layer, thereby eliminating the need for an insulating foam. Another K-type thermocouple was attached to the surface of the emitter to measure its temperature, $T_{abs}$, and estimate the blackbody emissive power. A convective shield was placed on top of the absorber to reduce convection losses from the surface. The tank was placed on a balance (A&D, GF-4000) that records the mass loss due to evaporation over time. A hygrometer (Onset, HOBO UX100) was kept next to the balance to measure ambient conditions (temperature, $T_\infty$, and relative humidity). The K-type thermocouples were connected to a data logger (Pico Technologies, USB TC-08) that records temperature as a function of time. A 35 mm aperture was placed above the absorber to limit extraneous light during experimental testing. As a first step, evaporation under dark conditions was measured for 30 minutes (solar simulator on with the shutter closed - dark evaporation). Next, the shutter was opened with the incident flux set to 1000 W m$^{-2}$, unless otherwise noted. Mass loss and temperature measurements were made for two hours under illumination, and values reported in the manuscript are averaged over four runs with the dark evaporation subtracted. Limitations of this prototype include a view factor less than unity (0.9), as well as convection losses from the top surface of the absorber.

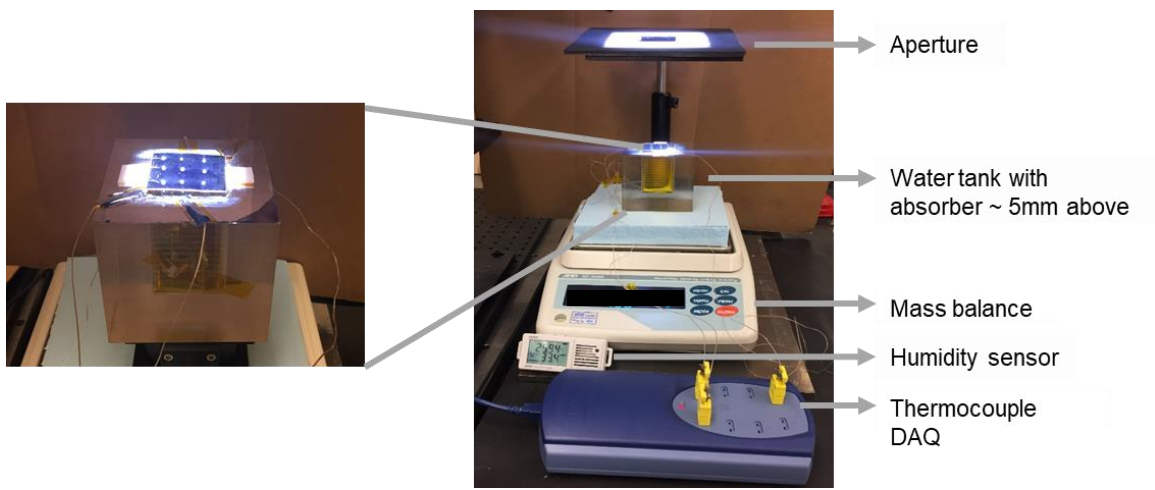

Fig. S4: Experimental setup for surface heating using a solar umbrella (selective solar absorber and black emitter), where the temperature profile and evaporation rate are monitored using thermocouples and a mass balance, respectively. Holes are drilled into the umbrella to serve as vapor escape pathways.

**Experimental Data**

The selective absorber shows a rapid response for conversion of solar energy into heat. Under a solar flux of 1000 W m$^{-2}$, the absorber reaches a peak operating temperature of 70 ± 2 °C in ~ three minutes owing to the use of a high thermal conductivity substrate (aluminum in this case). Figure S5 also shows the water evaporation rate under dark conditions, under direct illumination corresponding to volumetric heating, and using the solar umbrella corresponding to surface heating.

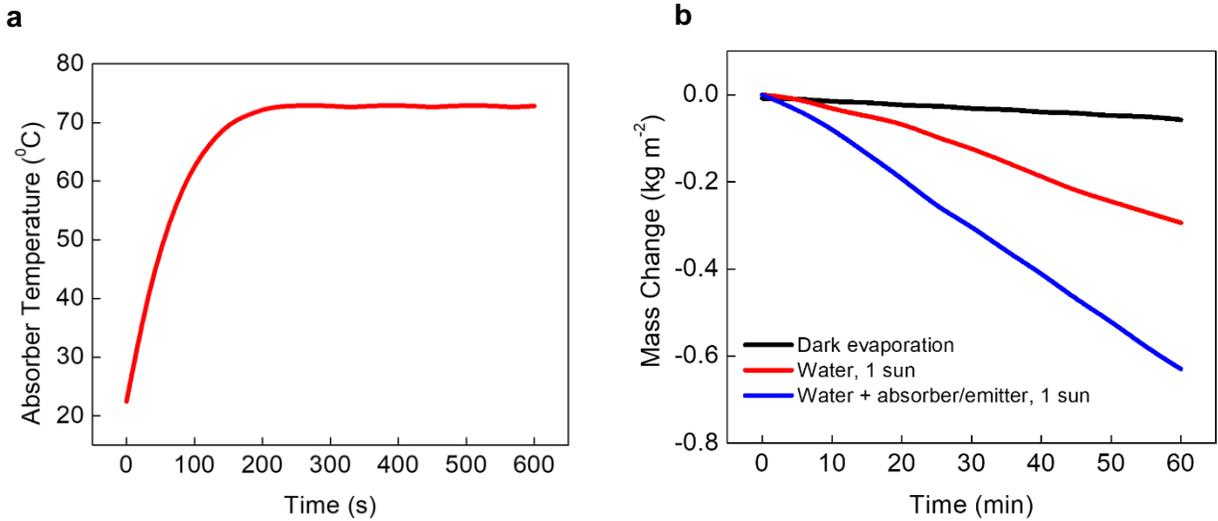

Fig. S5: (a) Temperature of the absorber-emitter when exposed to a solar flux showing a fast thermal transient response. (b) Mass change over time due to evaporation under dark conditions, and under one sun illumination with and without the absorber-emitter.

**Evaporation under Optical Concentration**

The lab-scale prototype was tested at different optical concentrations of 1, 3 and 5 suns. As expected, at higher solar fluxes, the absorber-emitter temperature increases, which in turn results in a higher emissive power on the water, thereby increasing the surface temperature. As an illustration, at 5000 W m$^{-2}$, the evaporation rate increases to 1.6 kg m$^{-2}$ h$^{-1}$ as the water surface temperature is 60 °C in one hour while the bulk remains at 30 °C. The absorber-emitter can also be used for generating hot vapor as the evaporated water quickly mixes with the air surrounding the hot absorber. A thermocouple was placed 2 mm below the absorber surface to measure the vapor temperature as it escapes from the holes. Significantly higher absorber temperatures can be achieved by using a better convective cover or an evacuated plate that eliminates convection losses from the hot absorber. Non-tracking wide-angle optical concentrators can provide a passive and cost-competitive approach to concentrate sunlight for evaporation.

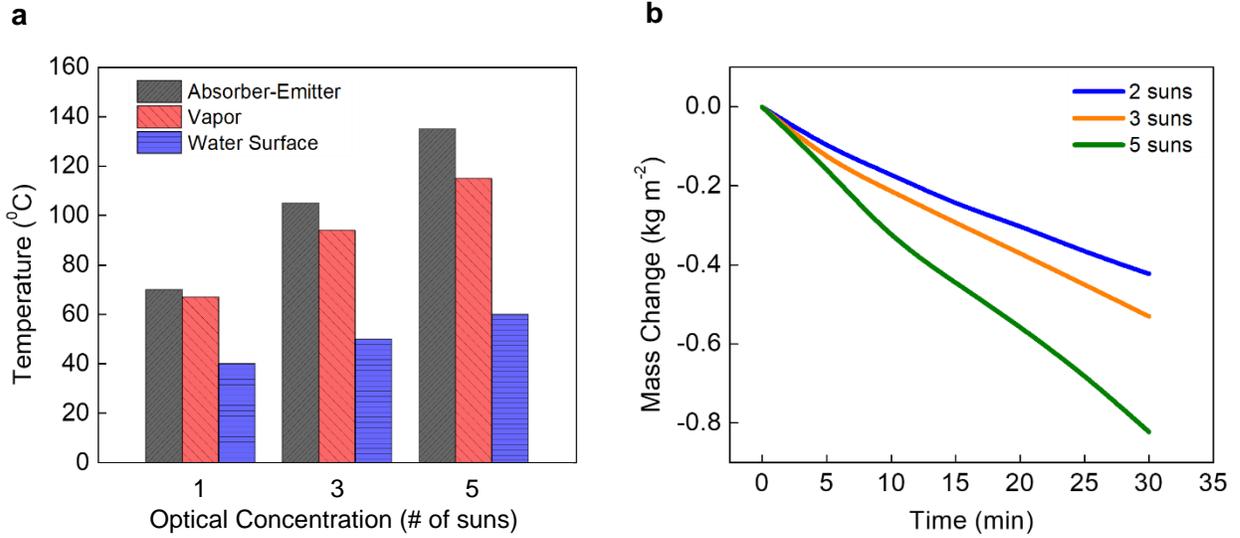

Fig. S6: (a) Temperatures of the absorber-emitter, water vapor (measured 2 mm below the absorber surface) and the water surface as a function of low optical concentrations. (b) Mass change over time due to evaporation under different solar fluxes.

## Heat and Mass Transfer Model

The overall system comprises two main sub-systems: the solar umbrella (absorber and emitter) and the water tank. The umbrella receives incoming solar flux ($Q_{solar}$), and absorbs a majority of it owing to a high solar absorptance of 0.95. As its temperature increases, 5% losses by thermal radiation ($Q_{rad}$) as well as convection losses at the top surface ($Q_{conv}$) occur to the ambient. Since the absorber and emitter coatings are thin layers on an aluminum substrate, the structure is isothermal, and the emitter radiates in the IR ($Q_{emit}$) to the water surface. Performing an energy balance on the umbrella,

$$Q_{solar} = Q_{conv} + Q_{rad} + Q_{emit}$$

Performing an energy balance on the water surface:

$$Q_{emit} = Q_{conv,w} + Q_{rad,w} + Q_{cond} + Q_{evap}$$

where $Q_{rad,w}$ and $Q_{conv,w}$ are the radiation and convection losses from water respectively (tank walls + water surface), $Q_{cond}$ is the conduction losses from the surface to the underlying bulk water and $Q_{evap}$ is the remaining energy that is available for evaporation. In the lab-scale prototype, due to the large view factor between the umbrella and water surface, radiation emitted by the water surface is largely absorbed by the emitter, which in turn emits back to the water surface. Less than 1% radiation loss occurs between the water surface and ambient, however the water tank walls exchange radiation with the ambient that constitutes thermal losses. Furthermore, since the umbrella surface is hotter than the water surface below, convection losses from the bottom surface of the umbrella and the water are negligible. In this case, heat transfer is by conduction through the air gap which is small compared to the radiation exchange between the two surfaces. However, convection losses from the water tank to the ambient exist and constitute thermal losses. Here:

$$Q_{rad,w} = F\varepsilon\sigma(T^4 - T_\infty^4)$$

$$Q_{conv,w} = h_{conv}(T - T_\infty)$$

$$Q_{cond} = -k\left(\frac{T_s - T_b}{L}\right)$$

$$Q_{evap} = \dot{m}h_{fg} = h_{evap}(T_s - T_\infty)$$

where $F$ is the view factor, $\varepsilon$ is the emittance, $T$ is the temperature of the surface, $T_\infty$ is the ambient temperature, $h_{conv}$ is the convective heat transfer coefficient, $k$ is the thermal conductivity of water, $T_s$ is the water surface temperature, $T_b$ is the bulk water temperature and $h_{evap}$ is the evaporation heat transfer coefficient. $h_{evap}$ is calculated as an average using $T_s$ and $\dot{m}$ values at different optical concentrations shown in Fig. S6 to obtain a value of 28 W/m²-K.

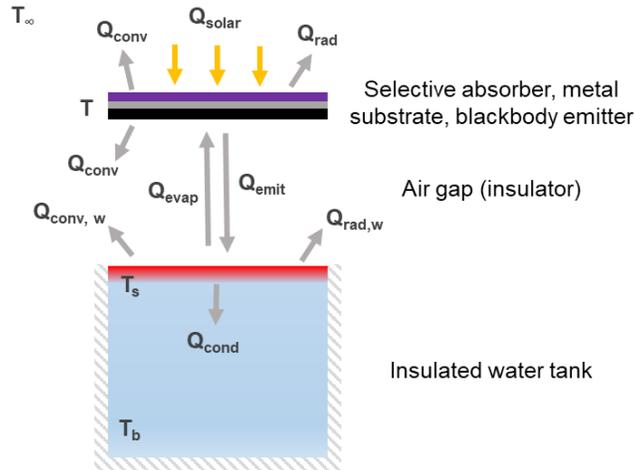

Fig. S7: Energy balance and heat transfer modes for the solar umbrella and water system.

**COMSOL Simulation**

The lab-scale setup is modeled using COMSOL Multiphysics software and the thermal losses are estimated. Water in the acrylic tank is modeled with the following boundary conditions: convection on all side walls = 5 W/m²-K, radiation from all side walls and water surface with emissivity = 0.95, thermal insulation on bottom surface, evaporation coefficient of 28 W/m²-K on water surface and a boundary heat flux of 680 W/m² on the water surface (this corresponds to the radiation from the emitter at 72 °C under 1 sun accounting for its emissivity and the view factor). The initial water temperature is set at 22.5 °C and the ambient temperature is 25 °C. Spectral radiation in participating media is used to model the high absorption coefficient of water in mid-IR (peak wavelength ~8 μm with absorption coefficient $10^4$ m⁻¹), resulting in heat localization at the surface. The absorber is modeled with a spectrally selective emissivity (absorptance of 0.95 for solar wavelengths and thermal emittance of 0.04 at wavelengths larger than 2.5 μm). The black paint emitter surface is modeled with an emittance of 0.94.

For the evaporation pond simulations, the geometry is modified to a surface with 1 m radius and a 0.5 m depth. Owing to the large surface area, convection and radiation losses from the sides of the system are negligible compared to the lab-scale prototype. Furthermore, for a large pond, the view factor between the umbrella and water surface is 1 (*i.e.,* all the radiation leaving the umbrella is intercepted by the water surface) so a boundary heat flux of 750 W/m² is used at the water surface.

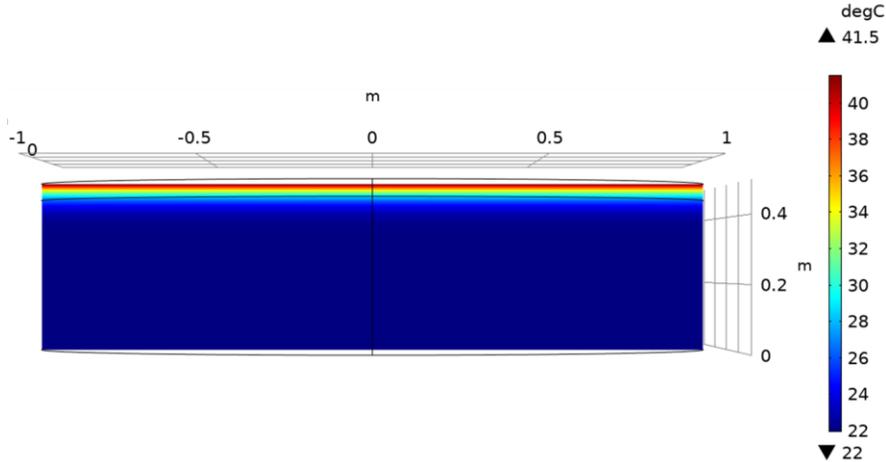

Fig. S8: COMSOL simulation of an evaporation pond showing heat localization at the surface when using a photo-thermal umbrella.